\documentclass{article}

\usepackage[a4paper]{geometry} 

\usepackage{natbib}
\usepackage{graphicx} 
\usepackage{url} 
\usepackage{amsmath}
\usepackage{amssymb}
\usepackage[table]{xcolor}
\usepackage{comment}
\definecolor{light-gray}{gray}{0.9}

\usepackage{etoolbox}
\makeatletter
\patchcmd{\@makecaption}
{\parbox}
{\advance\@tempdima-\fontdimen2} 
{}{}
\makeatother

\begin{document}
\title{On the dependence in football match outcomes:
	traditional model assumptions and an alternative proposal}
\author{Marco Petretta, Lorenzo Schiavon$^*$, Jacopo Diquigiovanni\\
\small{$^*$Department of Statistical Sciences, University of Padova, Padova, 35121, 
Italy.}
}

\maketitle

\begin{abstract}
The approaches routinely used to model the outcomes of football matches are characterised by strong assumptions about the dependence between the number of goals scored by the two competing teams and their marginal distribution. In this work, we argue that the assumptions traditionally made are not always based on solid arguments. Although most of these assumptions have been relaxed in the recent literature, the model introduced by Dixon and Coles in 1997 still represents a point of reference in the betting industry. While maintaining its conceptual simplicity, we propose a modification of the dependence structure. A real data application suggests that our model, named Mar-Co, outperforms the Dixon and Coles one in several betting scenarios, and parameter interpretation provides key insights on league dynamics.	
\end{abstract}
\textit{Betting; Dixon and Coles model; Football prediction; Marginal distribution; Poisson distribution; Under/Over}

\section{Introduction}
\label{intro}

Modelling match outcomes in association football (referred to simply as "football" hereafter) undoubtedly represents an element of primary interest in the field of sports analysis. In order to do this, two different but interconnected strategies can be considered: the results-based (or direct) approach and the goals-based (or indirect) approach. Given a specific match between two competing teams, the former focuses on modelling the categorical ordinal variable taking the three possible result values (home win-draw-away win) typically through a regression model in which the probabilities of the three final outcomes are estimated on the basis of some external variables \citep[see, e.g.,][]{koning2000balance, goddard2004forecasting, schauberger2018predicting, carpita2019exploring, groll2019hybrid}. Instead, the goals-based approach considers a broader framework in which the purpose is to model the number of goals scored by the two teams during that specific match.
Since estimating the probability of each possible combination of home goal and away goal allows the estimate of the probability of home win-draw-away win to be consequently obtained, the two approaches are nested. The goals-based approach, being more general, has also some intriguing practical consequences, such as allowing types of bet different from traditional 1-X-2 (e.g. Under/Over). In addition, the difference in terms of performance between the two strategies is investigated by \citet{goddard2005regression} and \citet{koopman2019forecasting}: whereas the former highlights no relevant differences between them, the latter finds evidence that the goals-based approach provides more precise forecasts. In light of all these reasons, in the rest of the paper we will focus on the goals-based strategy.

The first articles in this field are \citet{moroneyj} and  \citet{reep1971skill}, in which the Poisson distribution and the negative binomial distribution are proposed to model the aggregated number of goals scored per game. To the best of our knowledge, \citet{maher1982modelling} represents the first work aimed at modelling the number of goals scored by individual teams: specifically, the number of goals scored by the home team and the away team defines two independent Poisson random variables whose parameters depend on the attack and defence skills of the two teams. In the same paper, the author also carries out a bivariate Poisson model which keeps the two marginal distributions unchanged, driven by the fact that the initial proposal tends to underestimate the proportion of draws. In the nineteen-nineties two different models were proposed in the same year:  \citet{lee1997modeling} keeps the general structure of  \citet{maher1982modelling} assuming independence between the two Poisson random variables, whereas the article of \citet{dixon1997modelling} moves from the pioneering work of Maher by introducing some crucial innovations. First of all, the authors introduce a specific dependence structure by specifying a parameter $\rho$ which allows the joint probability to be different from the product of the marginal probabilities. Secondly, they include a weighting function $\phi$ which down-weights old matches in the likelihood in order to obtain estimates of the parameters that are mainly based on recent performances of the teams. In order to improve the aspect just mentioned, \citet{rue2000prediction} propose a Bayesian dynamic generalized linear model leaving the attack and defence parameters free to change randomly over time, while the aforementioned bivariate Poisson model is reconsidered and extended by \citet{karlis2003analysis}. A completely different approach is the one developed by \citet{baio2010bayesian}: in this case, their Bayesian hierarchical model considers two conditionally independent Poisson random variables for the numbers of goal scored,  but the dependence is introduced through a careful choice of the  hyper-parameters. A recent development of this approach is given by \citet{egidi2018combining}, where the betting odds are included in the model specification together with other modifications. \citet{owen2011dynamic} implements a dynamic generalized linear model whose evolution component is specified as a random walk for the attack and defence parameters, whereas \citet{mchale2011modelling} use copulas to allow dependence between the two Poisson random variables under the assumption that the dependence parameter can be expressed as  a  linear function of the rank difference of the two teams. As argued by the two authors, the copula model represents a more flexible solution than the bivariate Poisson since it also permits negative correlation. Another interesting proposal able to capture the main features mentioned so far is the one provided by \citet{koopman2015dynamic}, where a non-Gaussian state space model assumes a bivariate Poisson distribution whose attack and defence parameters are allowed to vary stochastically over time. 

Taking into account the variety of approaches presented in the literature, it is evident that the definition of a proper dependence structure between the goals scored by the two teams represents an essential issue in modelling the final outcome of a football game. The aim of this work is to enrich the current literature 
presenting a discussion of the common dependence assumptions of the models for football match outcomes and introducing an innovative method able to balance flexibility and conceptual simplicity. 
Specifically, moving from the Dixon and Coles approach, which still represents a point of reference in the betting industry \citep[see, e.g., the methodology used at][]{mercurius} due to its conceptual simplicity, we introduce a more comprehensive dependence structure to improve the overall forecasting performance, while maintaining the easiness in the parameter interpretation and data generator process definition. In order to do that, we define the joint probability mass function describing the number of goals scored by the two teams by means of carefully chosen marginal probability mass functions (Mar-) and conditional probability mass functions (-Co). In so doing, the resulting model (Mar-Co) still manages the dependence between the home and the away goals by using an univariate parameter as in the Dixon and Coles approach.

The article is organised as follows: in Section \ref{sec::dependence} a broad discussion about the aforementioned dependence structure (focusing in particular on the one proposed by Dixon and Coles) is proposed; we present our adjustments to the Dixon and Coles model in Section \ref{sec::mar-co_mod}, whereas in Section \ref{sec::results} an application compares our proposal with the Dixon and Coles approach. Finally, Section \ref{sec::conclusion} provides some concluding remarks.

\section{The dependence assumption}
\label{sec::dependence}

\subsection{On the existence of dependence}
\label{sec::exis_dep}

Differently from the results-based strategy, the goals-based one requires a careful management of the dependence between the number of goals scored by the two competing teams, a notoriously tough task \citep{karlis2009bayesian}. Indeed, different studies draw very different conclusions depending on the championship, period of time and statistical tool considered. For example, \citet{karlis2000modelling} highlight a small positive correlation combining the evidence  from 24 championships of different European countries by means of the method proposed by \citet{hasselblad199421}. In \citet{mchale2007modelling}, the correlations computed on the English Premier League data from August 2003 to March 2006 suggest that, in contrast to shots, the goals scored by the two teams show only slight positive or no correlation, whereas the work of \citet{mchale2011modelling} shows statistically significant negative correlation considering matches between national teams.

In this regard, a fundamental aspect needs to be clarified. Let $x$ be the number of goals scored by the home team in a generic match of a given championship and let $y$ be the number of goals scored by the opposing team (i.e. the away team). Similarly, let us define $x_{ij}$ and $y_{ij}$ the number of goals scored by teams $i$ and $j$ respectively in the specific match in which $i$ plays at home and $j$ plays away. Variables $x$ and $y$ are different random objects from $x_{ij}$ and $y_{ij}$ (although they are interconnected somehow), and so the development of an empirical study assessing the dependence between $x$ and $y$ is of limited practical use since the purpose is to model the team-specific variables $x_{ij}$ and $y_{ij}$. For instance, one is justified in expecting a team scoring many goals during a game to concede a low number of goals (i.e. negative correlation between $x$ and $y$) since in that specific occasion it totally outclasses the opposing team. On the other hand, a different dynamic can occur when two specific teams - characterised by specific attack and defence skills - play against each other: for example, a strong team playing against a weak one may score less goals than those expected since it mainly focuses on conserving energy for future more challenging matches. A discussion on the topic is also provided by \citet{dixon1998birth} and \citet{rue2000prediction}. 

However, the crucial empirical study of the dependence between $x_{ij}$ and $y_{ij}$ is practically unfeasible since each couple of teams plays against just twice during a season (and only once team $i$ plays at home and team $j$ away). Considering more than one season represents an intuitive solution, but it seems to be unsuitable since the strength of a team can vary widely from season to season (due to newly signed players, the sacking of a manager, etc\dots). In view of this, in the rest of Section \ref{sec::exis_dep} we will focus on the dependence between the aggregated variables $x$ and $y$ in order to have an indication about the general behaviour, but all the results provided should be approached with particular caution in light of the considerations just expressed.

In order to investigate the dependence, the data we consider---obtained from \url{http://www.football-data.co.uk/}---refers to the 9130 matches of the English Premier League, the French Ligue 1, the German Bundesliga, the Italian Serie A and the Spanish La Liga  played between the 2014-2015 season and the 2018-2019 season. A first, fundamental study concerns the existence of the most familiar kind of dependence---i.e. correlation---between $x$ and $y$. To do so, we consider a bootstrap test aimed at evaluating the Pearson correlation coefficient with the samples drawn under the hypothesis of independence. 
One may argue that Pearson correlation is specific to investigate linear correlation and it could be not suitable with count data with possible zero inflation presence. However, testing the Pearson coefficient value is sufficient to have indication against independence if linear correlation is observed in the data.
The observed value of the Pearson correlation coefficient (computed considering all the 9130 matches) is $-0.085$ and the p-value is $<0.001$. The same strong evidence against the hypothesis of independence can be found considering each championship separately, with the only small exception of the Italian Serie A, whose p-value is equal to $0.026$. This result confirms the evidence provided by \citet{mchale2011modelling} and consequently the importance of a proper modelling of the dependence between the number of goals scored during a match. 

\subsection{The Dixon and Coles dependence structure}
\label{sec:dix_col_dep_str}

Among the various goals-based approaches proposed in this field, the model introduced by \citet{dixon1997modelling} represents one of the most famous, innovative and performing. Since the model proposed in this paper moves from it, an overview of the Dixon and Coles approach is provided as follows. 

Let $x_{ij}^k$ be the number of goals scored by the home team $i$ against the away team $j$ in match $k$, $y_{ij}^k$ be the number of goals scored by the away team $j$ on the same occasion and let $m$ be the total number of teams considered. In order not to overcomplicate the notation, hereafter $x_k$ ($y_k$ respectively) will be used instead of $x_{ij}^k$ ($y_{ij}^k$ respectively) with the home team $i$ and the away team $j$ implicitly assigned to every match $k$.
First of all, the two authors define $x_k$ and $y_k$ as follows:
\begin{eqnarray*}
x_k &\sim& \text{Pois}(\lambda_k),  \label{eq:mar-pois} \\
y_k &\sim& \text{Pois}(\mu_k), \nonumber \\
\log(\lambda_k) &=& \gamma + \alpha_{i(k)} + \beta_{j(k)}, \nonumber\\ 
\log(\mu_k) &=&\alpha_{j(k)} + \beta_{i(k)},  \nonumber
\end{eqnarray*}
with $\text{Pois}(\lambda)$ denoting a Poisson distribution with mean $\lambda$,  $i(k)$ and $j(k)$ indices which identify the home and away teams playing match $k$, and where $\alpha$, $\beta$, $\gamma$ are the attack, defence and home effect parameter, respectively.
By considering the matches chronologically, the two authors divide the seasons into a series of half-weekly time points and construct the following function for each time point $t$:
\begin{align}
\label{eq::pseudoveros_DC}
\mathcal{L}_t(\alpha_i,\beta_i,\rho,\gamma; i=1,\dots,m)&=\prod_{k \in A_t} \biggr\{ \tau_{\lambda_k, \mu_k}(x_k, y_k) \, e^{-\lambda_k} \, \lambda_k^{x_k} \, e^{-\mu_k} \, \mu_k^{y_k} \biggr\} ^{e^{-\xi \, (t-t_k)}},  \\
A_t &= \{ k: t_k < t \}, \nonumber
\end{align}
with $t_k$ the time that game $k$ is played,  $\tau_{\lambda_k,\mu_k}(x,y)$ the function depending on parameter $\rho$ which manages the dependence between $x_k$ and $y_k$, and $\xi \geq 0$ the parameter that regulates the down-weighting of old matches. Consistently with the notation of the original paper, the constraint $\sum_{i=1}^m \alpha_i=m$ is included for identifiability.

Hence, the two authors obtain the estimates of the parameters by numerically maximising the function in Equation (\ref{eq::pseudoveros_DC}) at each time point $t$ after choosing $\xi$. 
The choice of $\xi$ is particularly tough because Equation (\ref{eq::pseudoveros_DC}) defines a sequence of non-independent functions that makes difficult to obtain the value of $\xi$ that maximises the overall predictive capability of the model. In order to overcome this problem, the two authors focus on the prediction of match outcomes rather than match scores and define the value of $\xi$ as the value maximising
\begin{equation*}
S(\xi)=\sum_{k} \left(\delta_{k}^{H} \log p_{k}^{H}+\delta_{k}^{D} \log p_{k}^{D}+\delta_{k}^{A} \log p_{k}^{A}\right)
\end{equation*}
with 
\begin{equation*}
p_{k}^{r}=\sum_{h, a \in B_{r}} \operatorname{pr}\left(x_{k}=h, y_{k}=a\right),
\end{equation*}
which is implicitly a function of $\xi$ since the score probabilities $\operatorname{pr}\left(x_{k}=h, y_{k}=a\right)$ are estimated from the maximisation of the function in (\ref{eq::pseudoveros_DC}) at $t_k$ with weighting parameter set at $\xi$ and with $r=\{H,D,A\}$, $B_{H}=\{(h, a): h>a\}, B_{D}=\{(h, a): h=a\}, B_{A}=\{(h, a): h<a\}$, and  $\delta_{k}^{H}$ ($\delta_{k}^{D}$, $\delta_{k}^{A}$ respectively) the delta function equal to 1 if the final result of game $k$ is a home win (draw, away win respectively).

The last, fundamental aspect introduced by Dixon and Coles concerns the description of the dependence structure, that is defined by means of the function 
\begin{equation}
\tau_{\lambda_k,\mu_k}(x,y)=\begin{cases} 
1-\lambda_k \mu_k \rho & \text{if $x=0$, $y=0$} \\ 
1+\lambda_k \rho & \text{if $x=0$, $y=1$} \\ 
1+\mu_k \rho & \text{if $x=1$, $y=0$} \\ 
1- \rho & \text{if $x=1$, $y=1$} \\ 
1 & \text{otherwise} 
 \end{cases}
\label{eq::function_tau}
\end{equation}
subject to 
\begin{equation*}
\max \biggr(-\frac{1}{\lambda_k},-\frac{1}{\mu_k}\biggr) \leq \rho \leq \min\biggr( \frac{1}{\lambda_k \mu_k},1\biggr).
\end{equation*}
By using function (\ref{eq::function_tau}), the marginal distributions of $x_k$ and $y_k$  are Poisson with means $\lambda_k$ and $\mu_k$ respectively and the independence between $x_k$ and $y_k$ is obtained when $\rho=0$. 

\begin{figure}
    \centering
    \makebox{\includegraphics[width=0.99\textwidth]{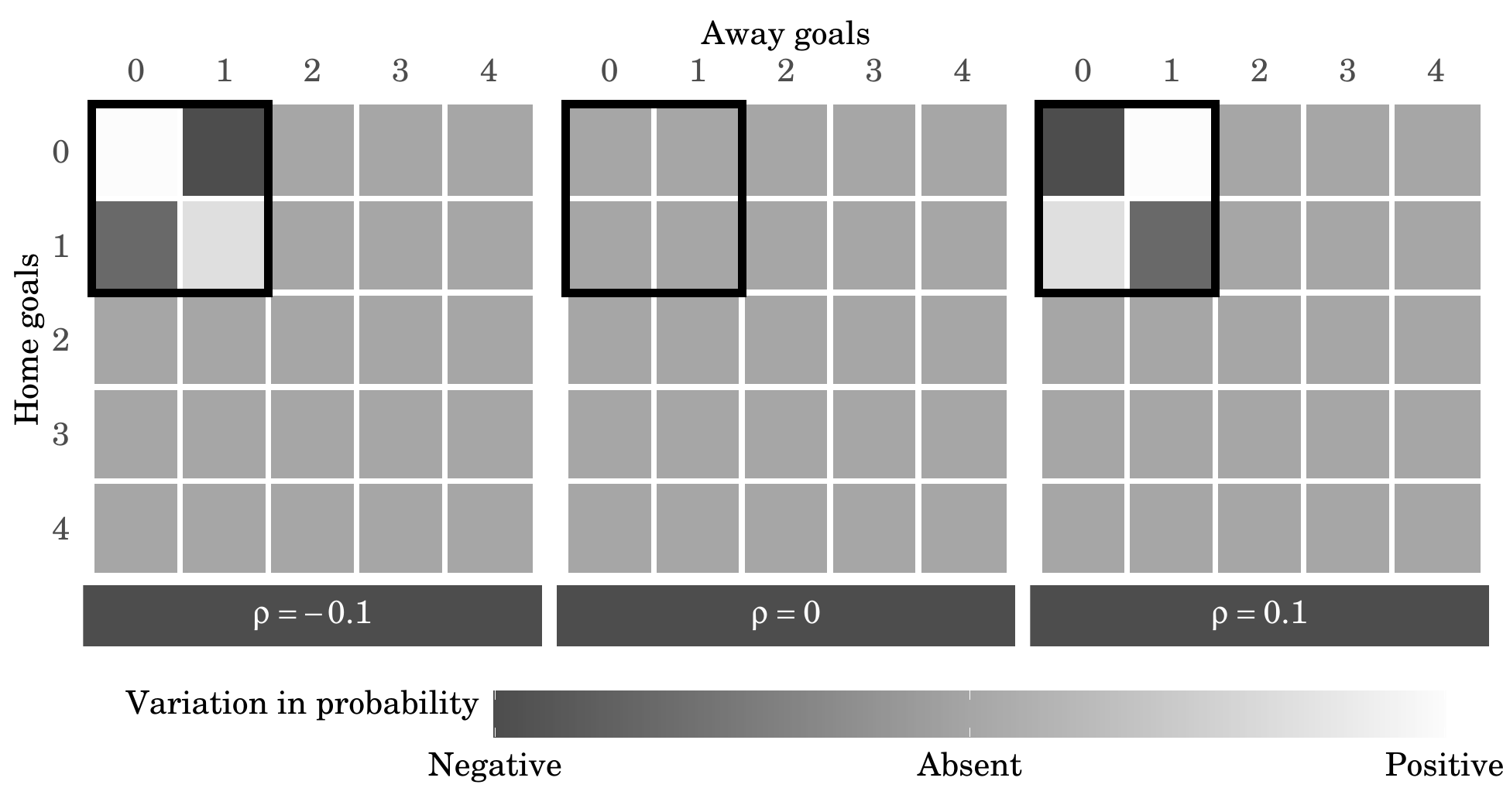}}
    \caption{\label{fig:dc}Each panel shows, under the Dixon and Coles model, the difference between the probability of the exact outcomes in match $k$, given a certain value of $\rho$ (equal to -0.1 on the left, 0 in the middle and 0.1 on the right), and the probability of the exact outcomes in match $k$ in case of independence (i.e. $\rho=0$). The marginal means $\lambda_k$ and $\mu_k$ are set equal to 1 and 1.5, respectively. }
\end{figure}

Figure \ref{fig:dc} provides an intuitive representation of the dependence structure, highlighting the difference between the joint probability mass functions of the goals scored assuming independence and dependence.
The only outcomes affected by a change in probability, when $\rho$ varies, are those in the black boxes of the picture. This means that the probability that one team scores at least two goals does not depend on the number of goals scored by the opposing team in the same match, since
$$
\sum_{h=0}^1 \sum_{a=0}^1 \text{pr}(x_k=h, y_k=a) = \sum_{h=0}^1 \sum_{a=0}^1 \text{pr}(x_k=h) \text{pr}(y_k=a)
$$
holds.
In other terms, the Dixon and Coles model is based on the strong assumption that the number of goals scored by the home team is independent of the number of goals scored by the away team, conditionally on observing an outcome not included in the set $\{(0,0), (1,0), (0,1), (1,1)\}$. 
Looking at the 9130 matches played during the period from August 2014 to May 2019 in the top five European championships, this assumption seems no longer reasonable, as suggested by testing it as null hypothesis in the following bootstrap procedure.
Let $\tilde{x}_k$ and $\tilde{y}_k$ denote the home and away number of goals considering scoring either zero or one goal as unique event---i.e. $\tilde{x}_k=\max(1,x_k)$, $\tilde{y}_k=\max(1,y_k)$. Hence, under the null hypothesis, the variables $\tilde{x}_k$ and $\tilde{x}_k$ are independent. To perform the test, we generated 1000 data sets with $\tilde{x}_k$ and $\tilde{y}_k$ independently re-sampled and we computed the sample correlation between the two variables in each data set. The negative correlation computed on the observed data set lies in the left tail of the bootstrap correlation distribution, encouraging the rejection of the null hypothesis with a p-value $<0.001$.
\begin{table}
	\caption{\label{tab:rep_dc} Ratios between the observed frequencies of 25 possible match outcomes and the joint probability function obtained by multiplying the observed marginal probability functions of home and away scores. Bootstrap standard errors are reported in parentheses, while grey cells indicate ratios that are significantly different from 100 at level $0.05$. To facilitate the reading, the numbers are multiplied by 100.}
	\fbox{%
	\begin{tabular}{crrrrr}
	Home& \multicolumn{5}{c}{Away goals}\\
	goals & 0 & 1  & 2 & 3 & 4 \\
	\hline
	0 & 99.47 (2.25) & \cellcolor{light-gray} 92.95 (2.02)  & 100.68  (3.10) & \cellcolor{light-gray} 114.96  (5.13) & \cellcolor{light-gray} 139.57  (9.53) \\
	1 & \cellcolor{light-gray} 94.46 (1.77) & 102.11 (1.65) & 102.75  (2.35) & 103.21  (3.91) & 107.72  (6.86)\\
	2 & 99.01 (2.10) & 103.40  (1.98) & 102.69  (2.92) & \cellcolor{light-gray}  89.23  (4.55) & \cellcolor{light-gray} 84.39  (7.75) \\
	3 & \cellcolor{light-gray} 108.13 (3.18) & 101.23  (2.95) & \cellcolor{light-gray} 91.24  (4.12) &  99.11  (7.13) & \cellcolor{light-gray} 59.27 (10.11)\\
	4 & \cellcolor{light-gray} 121.20 (5.29) & 98.78  (4.73) & 88.63  (6.65) & \cellcolor{light-gray} 68.19  (9.49) & \cellcolor{light-gray} 52.58 (14.97) 
	\end{tabular}}
\end{table}
As a further evidence against the Dixon and Coles dependence structure, we replicated on the above mentioned recent data the study reported in Section 3 of \citet{dixon1997modelling}. Table \ref{tab:rep_dc} reports the ratio $f_{x,y}(x_k,y_k)/\{f_x(x_k)f_y(y_k)\}$, where $f_{x,y}(x_k,y_k)$, $f_x(x_k)$, and $f_y(y_k)$ are the joint and marginal observed probability mass functions for the home and away number of goals, respectively. We notice that nowadays, unlike what observed by Dixon and Coles at the time, the frequency $f_{x,y}(x_k,y_k)$ of several match outcomes is significantly different from the product of the marginal frequencies of goals scored by home and away teams $f_x(x_k)f_y(y_k)$.

These empirical results, together with the evidence  that a kind of dependence between the home and away goals exists, suggest that a new, more complex dependence structure should be considered.
Furthermore, the Dixon and Coles model aims to properly estimate the probabilities of home win-draw-away win, which are directly influenced by the dependence structure proposed, since each of these three probabilities changes according to $\rho$. Vice versa, such dependence structure cannot adjust the probability of events defined by other types of bet, as we will see in details in Section \ref{sec::results}. 
Motivated by these considerations, in the next section, starting from the Dixon and Coles model, we propose a new model characterised by a more general dependence structure, such that the induced joint probability mass function is different from the product of the marginal probability mass functions for each possible match outcome.

\section{The Mar-Co model}
\label{sec::mar-co_mod}
Inspired by \citet{berkhout2004bivariate} and moving from the considerations discussed in the previous section, we propose a new model which differs from the Dixon and Coles one by the choice of the dependence structure between $x_k$ and $y_k$. 

Given a certain marginal distribution of $x_k$, an intuitive approach to allow the specification of a sufficiently general dependence structure consists in modelling the conditional distribution $y_k \mid x_k$. 
By indicating, with a slight abuse of notation, $F_{\mu_k}(x)$ 
as the cumulative distribution function evaluated at $x$ of a Poisson random variable with mean $\mu_k$---i.e. $F_{\mu_k}(x)=\exp\{-\mu_k\} \sum_{i=0}^x( \mu_k^i/i!) \; \forall \, x \in \mathbb{N}_{0}$---
we specify 
\begin{eqnarray*}
	(y_{k} \mid x_{k} = h) &\sim& \text{Pois}\{\psi_{y}(\mu_{k},h)\}, \qquad \text{with}\\
	\psi_{Y}(\mu_{k},h) &=&  \exp[\theta_{1} + \theta_{2}\log{\mu_{k}} + \theta_{3} \text{logit}\{F_{\mu_k}(h)\}], 
\end{eqnarray*}
where logit$(p)=\log\{p/(1-p)\}$ and $\boldsymbol{\theta}=(\theta_1, \theta_2, \theta_3)^\top \in \mathbb{R}^{3}$.
From an interpretative point of view, the mean of $y_k \mid x_k$ depends on the ability of the away team, characterized by $\mu_{k}$, to score no more goals than those scored in that occasion by the home team. 

As in the Dixon and Coles model, an univariate parameter ($\theta_3$ in this case) regulates the dependence between $x_k$ and $y_k$, where $\theta_{3} \neq 0$ implies dependence existence.
In addition, it is possible to obtain the above mentioned independence case corresponding to the Dixon and Coles model with $\rho=0$ by assuming $\boldsymbol{\theta}=(0,1,0)^\top$ and by modelling $x_k$ as the Poisson random variable defined in (\ref{eq:mar-pois}). 

Hence, assuming $x_k \sim \text{Pois}(\lambda_k)$, we can exploit the definition of joint probability mass function as product of conditional and marginal probability mass functions to define
\begin{equation}
\label{eq:P_A}
\text{pr}_A(x_k=h, y_k=a)=\frac{e^{-\psi_y(\mu_{k},h)} \psi_y(\mu_{k},h)^a}{a!} \frac{e^{-\lambda_{k}} \lambda_{k}^h}{h!}.
\end{equation}
On the other hand, if we follow symmetrical steps specifying
\begin{eqnarray*}
	y_k &\sim& \text{Pois}(\mu_k),\\
	(x_k\mid y_k =a) &\sim& \text{Pois}\{\psi_{x}(\lambda_{k},a)\},\\
	\psi_{x}(\lambda_{k},a) &=&  \exp[\theta_{1} + \theta_{2}\log{\lambda_{k}} + \theta_{3} \text{logit}\{F_{\lambda_k}(a)\}],
\end{eqnarray*}
we can define the symmetrical joint probability mass function
\begin{equation}
\label{eq:P_B}
\text{pr}_B(x_k=h, y_k=a)=\frac{e^{-\psi_x(\lambda_{k},a)} \psi_x(\lambda_{k},a)^h}{h!} \frac{e^{-\mu_{k}} \mu_{k}^a}{a!}.
\end{equation}
As we do not have formal reasons to favour one of the two specifications, we define the distribution of the joint outcome $(x_k, y_k)$ as an equally weighted mixture of the two distributions defined by $\text{pr}_A$ and $\text{pr}_B$. Therefore, under the Mar-Co model, the likelihood related to the $k$-th match can be expressed as 
\begin{eqnarray*}
\mathcal{L}_k^{M}(\alpha_i, \beta_i, \alpha_j, \beta_j, \gamma, \boldsymbol{\theta}) &\propto& \text{pr}_A(x_k=h, y_k=a; \alpha_i, \beta_i, \alpha_j, \beta_j, \boldsymbol{\theta}, \gamma) \\
 &\quad& + \text{pr}_B(x_k=h, y_k=a,\alpha_i, \beta_i, \alpha_j, \beta_j, \boldsymbol{\theta}, \gamma ). \nonumber
\label{eq:Lt} 
\end{eqnarray*}
As a merely football-based argument in favour of this model construction, 
the presence of two symmetrical data generation processes appears to be plausible, since one is justified in expecting sometimes the away team to react to the performance of the home team, sometimes vice versa.

\begin{figure}
    \centering
    \makebox{\includegraphics[width=0.99\textwidth]{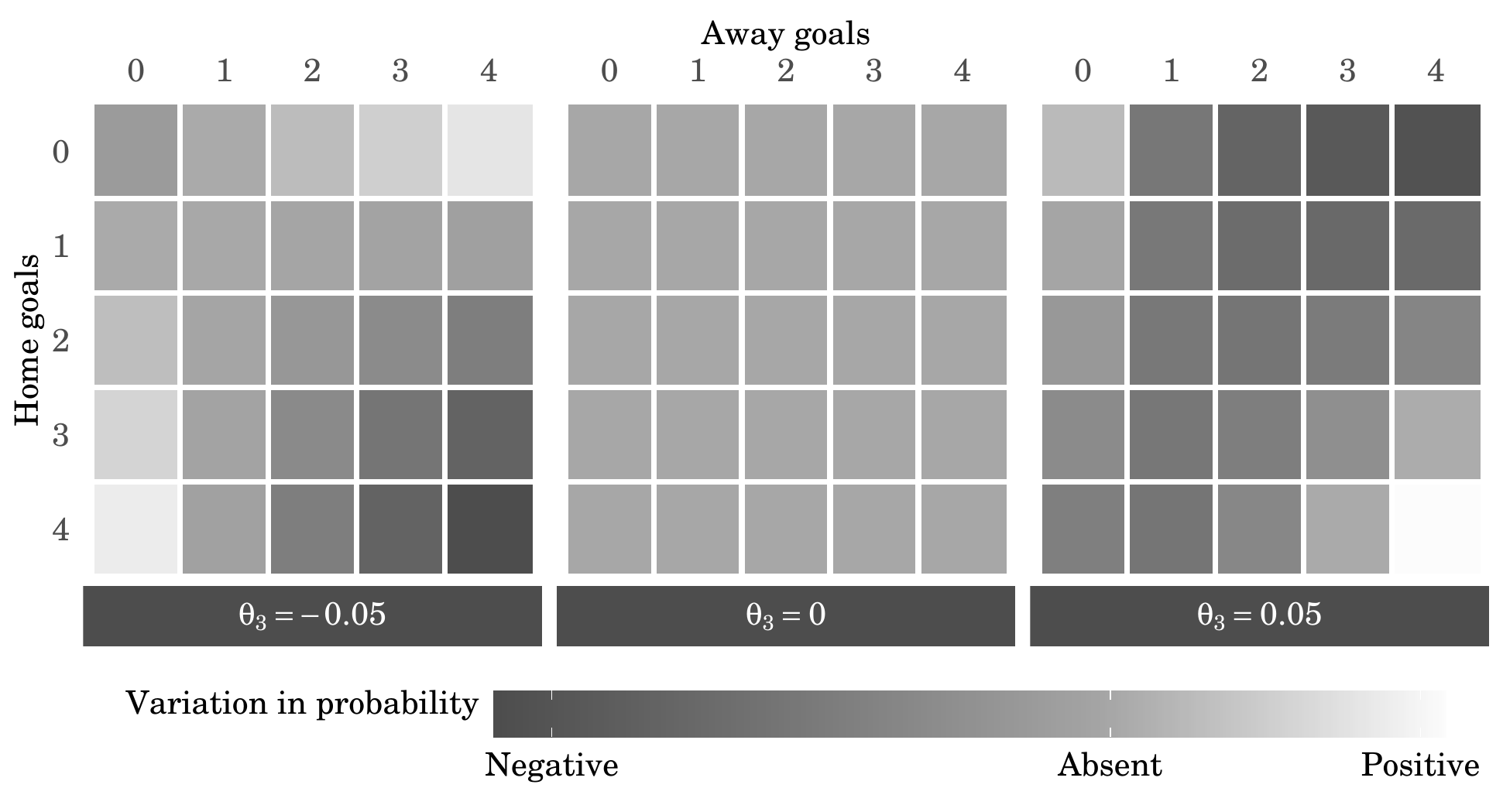}}
    \caption{\label{fig:marco} Each panel shows, under the Mar-Co model, the difference between the probability of the exact outcomes in match $k$, given a certain value of $\theta_3$ (equal to -0.05 on the left, 0 in the middle and 0.05 on the right), and the probability of the exact outcomes in match $k$ in case of independence (i.e. $\theta_3=0$). The marginal means $\lambda_k$ and $\mu_k$ are set equal to 1 and 1.5, respectively, while $\theta_1=0$ and $\theta_2=1$. }
\end{figure}
The key difference between the Mar-Co model and the Dixon and Coles one lies in the specification of a joint distribution that permits $x_k$ and $y_k$ to be dependent also conditionally on observing an outcome not included in the set $\{(0,0),(1,0),(0,1),(1,1)\}$. This aspect is clearly displayed in Figure \ref{fig:marco}, where the grey scale identifies the difference between the Mar-Co joint probability mass function considering three possible values of $\theta_3$ and the probability mass function obtained when $\theta_3=0$. A comparison of the left and right panels of Figure \ref{fig:marco} with the corresponding panels in Figure \ref{fig:dc} immediately highlights the discrepancy between the two models in terms of outcomes affected by the introduction of dependence. 
Formally, 
\begin{align*}
	\sum_{h =0}^1 \sum_{a=0}^1 \text{pr}(x_k=h, y_k=a) &=
	\sum_{h =0}^1 \sum_{a=0}^1 \frac{\text{pr}_A(x_k=h, y_k=a) + \text{pr}_B(x_k=h, y_k=a)}{2} \\
	&\neq \sum_{h =0}^1 \sum_{a=0}^1\text{pr}(x_k=h) \text{pr}(y_k=a),
\end{align*}
when $\theta_3 \neq 0$.
The parameter $\theta_3$ presents also an immediate interpretation, which can be easily described through well known team behaviours unlike the parameter $\rho$ in the Dixon-Coles models or other complicate dependence structures. Indeed, $\theta_3>0$ means that teams tend to adapt their attacking performance to the opponent performance such that several tight matches are generated. In other words, teams tend to increase their attacking level against well-performing teams and to lower it when big attacking efforts are not necessary to win. Vice versa, if $\theta_3<0$, teams tend to have better (worse) performance than their \textit{marginal} behaviour when they face teams playing poorly (very well). In fact, both these behaviours are likely and they generally depend on the football culture of a team or, more often, of a country, as will be discussed in Section \ref{sec::results}. 
Relying on distinct parameter estimation processes for the different leagues, one may interpret the estimates of $\theta_3$ as an average league measure of the competitive balance within the matches, with $\theta_3$ positive indicating a league where matches are often tighter than what expected by looking at the marginal strength of the teams involved. 
A careful analysis of the properties and a proper definition of $\theta_3$ as an index of competitiveness may lead to the inclusion of a further dimension in the framework of competitive balance measurements, a field that is gaining large attention from researchers in recent years \citep{manasis2014, manasis2022}.
Although the latter objective is beyond the aim of this work, the potentialities of our proposed dependence structure in terms of interpretation represent a further distinct and interesting trait with respect to other more complex models.

\begin{figure}
	\centering
	\includegraphics[width=0.99\textwidth]{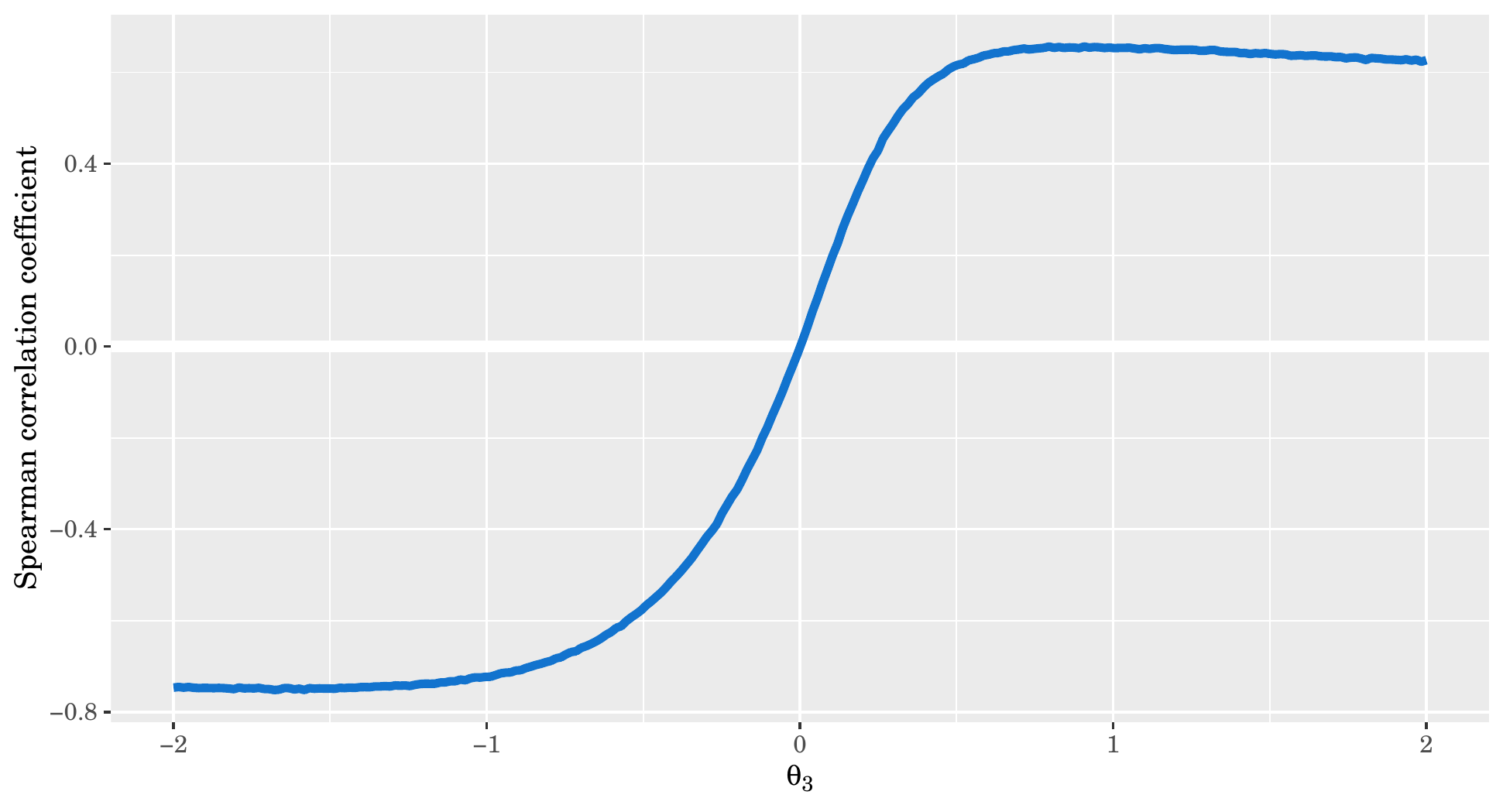}
	\caption{Monte Carlo estimates of the Spearman correlation coefficient at varying of $\theta_3$, with $\theta_{1}=0$, $\theta_{2}=1$, $\mu_k=1.75$, and $\lambda_{k}=1.25$. The estimates are based on $250$ computations of sample Spearman coefficient of $1000$-large  samples of possible match outcomes.}
	\label{fig:correlation}
\end{figure}

On the face of a more comprehensive dependence structure, the Mar-Co model does not present the desired properties in terms of marginal distributions that characterise the Dixon and Coles model.
Indeed, in case of dependence under the Mar-Co model, both the conditional and the marginal distributions of $x_k$ and $y_k$ are not Poisson, since each of these distributions can be described as the mixture between a Poisson and a non-Poisson distribution.
When $\theta_3 \neq 0$, the marginal distributions $x_k$ and $y_k$ are characterised by non finite moments, since they linearly depends on the non finite expected values $E([F_{\lambda_k}(y)/\{1-F_{\lambda_k}(y)\}]^{\theta_3})$ or $E([F_{\mu_k}(y)/\{1-F_{\mu_k}(x)\}]^{\theta_3})$, where $F_{\lambda_k}(y)$ and $F_{\mu_k}(x)$ are uniformly distributed in $(0,1)$. Although this fact is not appealing in terms of property demonstrations, it suggests a robust behaviour of the marginal distributions with respect to large results, due to the fat tails.
The non finite first moments of the marginal distributions does not allow one to analytically recover the correlation between home and away goals. Then, the induced correlation between $x_k$ and $y_k$ at varying of $\theta_3$ is reported in Figure \ref{fig:correlation} via Monte Carlo estimates of the Spearman correlation coefficient, which is robust to non finite moments. 

The lack of Poisson marginal distributions appears in contrast with one of the most widely accepted assumptions in the literature. As mentioned in the Introduction, several authors assume Poisson marginal distributions on the basis of the empirical behaviour of the marginal home and away number of goals in a championship \citep[see, e.g., ][]{dixon1997modelling}. 
Nevertheless, a parametric bootstrap test conducted on the same data presented in Section \ref{sec::exis_dep} questions the validity of this assumption, suggesting to reject the null hypothesis of Poisson-distributed variables $x$ and $y$.
In details, we compared the ratio between the sample variance and the sample mean---i.e. an estimate of the dispersion parameter---computed on the observed data with those computed on 1000 data sets drawn from the Poisson distribution with mean equal to the observed sample mean. In so doing, we observed a p-value $<0.001$ both for the home and the away number of goals.
By replicating the test on each championship separately, we obtained similar strong indications against the null hypothesis for each marginal distribution considered. In addition, the same evidence, although slighter, was obtained by performing an alternative parametric bootstrap test.
This later test is based again on 1000 data sets drawn from the Poisson distribution with mean equal to the observed sample mean and it considers as test statistic the Kullback-Leibler divergence between the empirical distribution of each sample and the theoretical Poisson distribution used to generate the samples.

Following the considerations discussed in Section \ref{sec::exis_dep}, a further crucial element should be taken into account. 
The studies of the marginal behaviour of the number of goals, such as that reported above, generally investigate the properties of the distributions of $x$ and $y$, i.e. the number of goals observed in a generic match of a championship, while the goals-based models usually make assumptions on the distributions of $x_{ij}$ and $y_{ij}$, i.e. the number of goals scored in a match between two specific teams $i$ and $j$. The article of \citet{karlis2000modelling} represents a progress in this discussion, since the authors analyse the dispersion parameter of the distribution of the number of goals in matches played by a given team $i$, i.e. $x_{i}$ and $y_i$, 
providing an evidence that we can expect to be more similar to that we could obtain by analysing the distribution of $x_{ij}$ and $y_{ij}$. The study, conducted on 456 teams in 24 different European leagues, suggests the presence of overdispersion, providing a further element against the hypothesis of Poisson-distributed variables $x_{ij}$ and $y_{ij}$.
Therefore, although we cannot conclude that $x_{ij}$ and $y_{ij}$ are undoubtedly non-Poisson random variables, it is hard to argue that the assumption of marginal Poisson distributions is usually a convenient choice in terms of modelling, rather than a consequence of well established behaviours. 
In view of this perspective, it could be worth having non-Poisson marginal distributions, if it allows to include a more flexible dependence structure able to improve the overall predictive capability of the model, as we try to demonstrate via an application study reported in the next section.

As regards the estimation process, let $\boldsymbol{\alpha}$ and $\boldsymbol{\beta}$ denote the vectors including the parameters $\alpha_i$ and $\beta_i$ for $i=1,\ldots,m$.
The estimates of the parameters at time point $t$ are obtained by maximising the function
\begin{equation}
	\mathcal{L}_t^{M}(\boldsymbol{\alpha},\boldsymbol{\beta},\gamma, \boldsymbol{\theta})=\prod_{k \in A_t} \biggr\{ \text{pr}_A(x_k, y_k; \boldsymbol{\alpha}, \boldsymbol{\beta}, \gamma, \boldsymbol{\theta}) + \text{pr}_B(x_k, y_k;\boldsymbol{\alpha}, \boldsymbol{\beta}, \gamma, \boldsymbol{\theta} )\biggr\}^{e^{-\xi \, (t-t_k)}}, \nonumber
	\label{eq::pseudoveros_M}
\end{equation}
where $\text{pr}_A$ and $\text{pr}_B$ refer to the functions in (\ref{eq:P_A}) and (\ref{eq:P_B}) respectively, while the set $A_t$ and the parameter $\xi$ are defined consistently with the notation of the Dixon and Coles model presented in Section \ref{sec:dix_col_dep_str}.
Given the shape of the mixture distribution, the presence of local modes cannot be excluded and, at the same time, the maximisation algorithms could encounter computational troubles in evaluating $\mathcal{L}_t^M$ at points that are far from the mode.
However, empirical results suggest that the maximisation procedure followed in this work generally leads to stable and satisfying results.

\section{An application to European leagues}
\label{sec::results}
In this section, a detailed comparison between the model presented in Section \ref{sec::mar-co_mod} and the Dixon and Coles one is provided. The two models are evaluated on a real data application which considers the  five most important European leagues (the English Premier League,  the French Ligue 1, the German Bundesliga, the Italian Serie A, the Spanish La Liga) and different betting types. Although sports betting is not the primary tool to assess the performance of a model, in this specific case it is particularly useful since it allows to compare the two models in different frameworks and to highlight strengths and weaknesses
of each approach.

First of all, we focus on the most famous betting type, i.e. Home Win-Draw-Away Win (known simply as 1-X-2).
In order to compare the two models, the first step is the choice of the parameter $\xi$, that is made for each combination of league and model separately by means of the procedure proposed by Dixon and Coles and described in Section \ref{sec:dix_col_dep_str}. In fact, consistently with the work of \citet{diquigiovanni2019analysis}, hereafter the time point $t$ will refer to the specific day of the year on which a given match takes place. This is due to the fact that nowadays, unlike when Dixon and Coles carried out the study, the teams play almost every day and so a more precise subdivision of the season is required. Specifically, the matches of three consecutive seasons (2012-2013, 2013-2014, 2014-2015) are used to choose $\xi$ and the probabilities of home win-draw-away win are not estimated for matches played between May and September in order to include promoted teams and to avoid misleading results due to lack of effort of some teams in the final part of the season. In so doing, the estimates of the probabilities are obtained starting from the first game played in October 2013 on the basis of all the information available at that time and  $S(\xi)$ is computed considering that game together with all the subsequent matches. 

Table \ref{tab:opt_xi} displays the values of $\xi$ maximising $S(\xi)$ based on a grid search, with $\xi>0$ indicating exponentially decreasing importance of a match over the time. 
\begin{table}
	\caption{\label{tab:opt_xi}Values of $\xi$ maximising $S(\xi)$ for each combination of league (columns) and model (rows).}
	\centering
	\fbox{%
	\begin{tabular}{lccccc}
	arg$\max_{\xi}$ $S(\xi)$ &England & France  & Germany & Italy & Spain \\
	\hline
	Dixon and Coles Model & 0 & 0 &0.0046&0.0053&0.0021 \\
	Mar-Co Model & 0 & 0 &0.0025&0.0045&0.0021 \\
	\end{tabular}}
\end{table}
The results seem  to show on the one hand no relevant dissimilarities between the two models, and on the other different dynamics in the leagues considered. 
Although it is not an aspect of primary interest, this evidence suggests a fascinating insight into European leagues: as the greater the value of $\xi$, the less importance is given to the oldest matches, then the current physical and psychological condition of the two competing teams seems to particularly affect the  outcome of a match in the German Bundesliga and the Italian Serie A. 

Once the value of $\xi$ is set, we use the Ranked Probability Score \citep[or RPS;][]{epstein1969scoring} to compare the performance of the two models. 
Despite the well known limitations of it \citep{wheatcroft2019evaluating}, RPS still represents one of the most famous and used scoring rules in this field due to its conceptual simplicity and easiness of implementation. 
Formally, the RPS related to a specific match and betting type is defined as
\begin{equation*}
 \frac{1}{r-1} \sum_{i=1}^{r-1}\left(\sum_{j=1}^{i}\left(\hat{p}_{j}-e_{j}\right)\right)^{2}
\end{equation*}
with $\hat{p}_j$ the estimate of the probability of the $j$-th possible betting outcome and $e_j=1$ if the outcome $j$ is observed and 0 otherwise. In the 1-X-2 framework, we therefore obtain: $r=3$, $\hat{p}_1$ ($\hat{p}_2$, respectively) equal to the estimate of the probability of home win (draw, respectively) and $e_1=1$ ($e_2=1$, respectively) if the final result of the match is a home win (draw, respectively) and 0 otherwise. 
A detailed description of the RPS and its properties can be found in \citet{constantinou2012solving} and references therein.
\begin{figure}
 \includegraphics[width=0.99\textwidth]{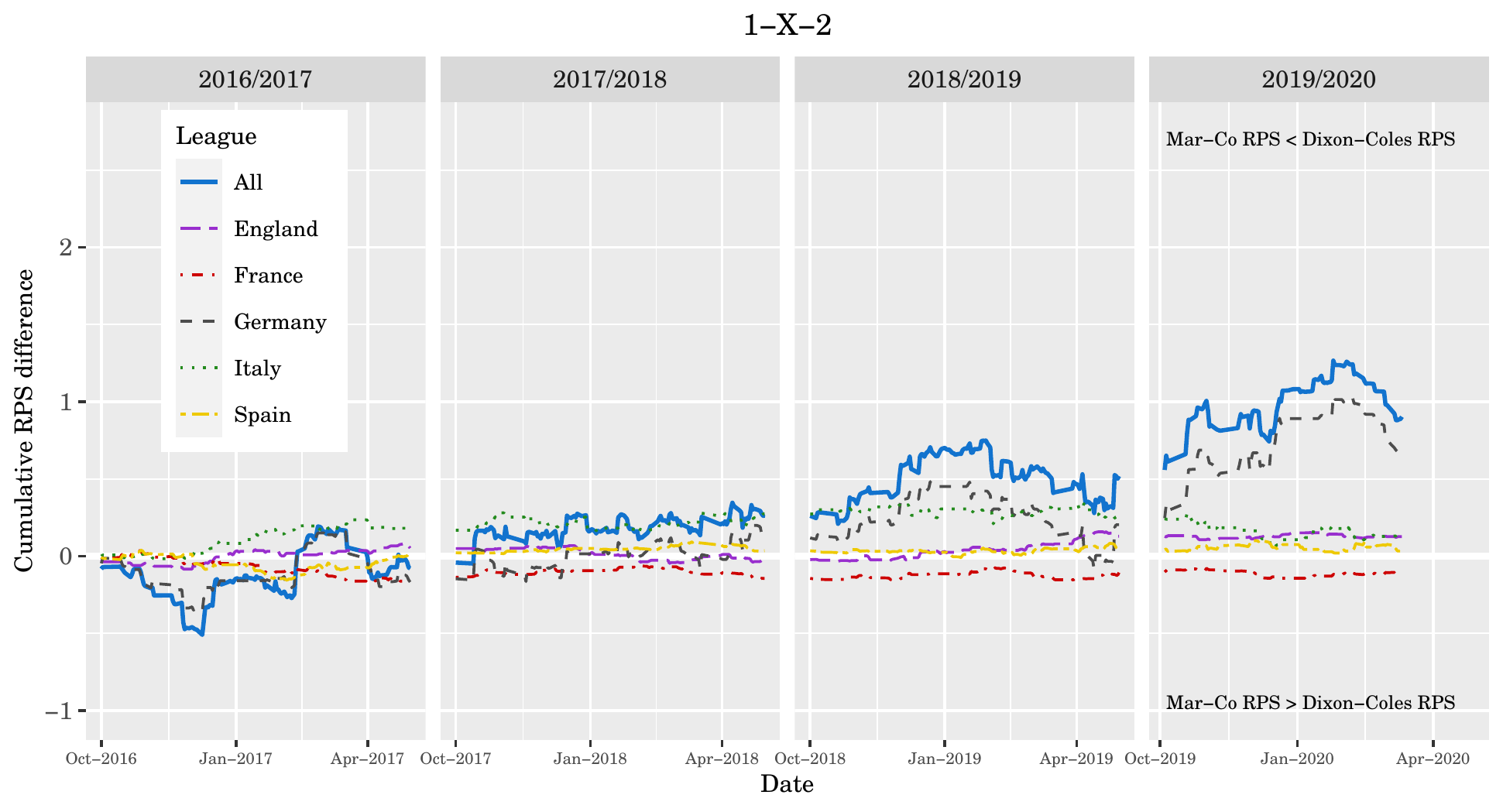}
\caption{Cumulative difference between the RPSs computed for the two competing models. Betting type: 1-X-2.}
\label{fig:1}       
\end{figure}
We compute the RPSs for the matches of the 2015-2016, 2016-2017, 2017-2018, 2018-2019, 2019-2020 (until interruption due to the COVID-19 outbreak) seasons.
 As for the choice of $\xi$, the first RPSs are obtained starting from the second season and they are computed only for matches played between October and April. 

Figure \ref{fig:1} shows the cumulative difference over time between the RPSs computed for the Dixon and Coles model and the RPSs computed for the Mar-Co model considering the matches of all the leagues 
(solid blue line) and considering each league separately (dashed coloured lines). Since a smaller RPS indicates better predictive performance, values 
above  the horizontal white line suggest that the Mar-Co model outperforms the Dixon and Coles one, and vice versa. The trend over time seems to be quite promising: after an initial period characterised by some fluctuations, the Mar-Co model provides more accurate predictions than those outputted by the Dixon and Coles one considering both all the leagues together and each league separately, with the sole exception of the French Ligue 1. Although the overall evidence seems to suggest that the modification introduced by the Mar-Co model can be used profitably in the long run, the analysis of Figure \ref{fig:1} should be accompanied by a study that verifies whether the two models are statistically different or not. To do that, for every match considered, the RPS 
computed for the Mar-Co model and the RPS 
computed for the Dixon and Coles one are switched with probability $0.5$ in order to obtain a new sample. By replicating this procedure $n_{b}$ times, it is possible to compute the $n_{b}$ mean differences of the reshuffled samples: in so doing, if the evidence in favour of the Mar-Co model is frequently stronger than the observed one, we conclude that the observed difference between the two models is mainly due to chance. By applying this procedure with $n_{b}=10000$, we obtain that 14\% of the time a stronger evidence is obtained: we conclude that, although it cannot be excluded that the two models are identical in terms of performance, the Mar-Co model seems to represent a promising modification of the starting model.

\begin{table}
	\caption{\label{tab:theta} Estimates $\hat{\theta}_3$ for each league and the related bootstrap $95\%$ confidence intervals (CI) based on $250$ replications.}
	\centering
	\fbox{%
		\begin{tabular}{lrl}
		Ligue &$\hat{\theta}_3$ & $95\%$ CI \\
		\hline
		England &  $-0.013$ &  $(-0.038, 0.015)$ \\
		France  & $-0.027$ & $(-0.051, 0.000)$  \\
		Germany &  $-0.078$ & $(-0.119, -0.040 )$  \\
		Italy &  $0.003$ & $(-0.044, 0.043 )$  \\
		Spain &   $0.015$ &  $(-0.011, 0.046)$ \\
	\end{tabular}}
\end{table}
Since the two models differ with regard to their dependence structure, parameter $\theta$ represents the key element in determining the satisfactory results obtained by the Mar-Co model. As a consequence, the estimate $\hat{\theta}_3$ is a quantity of deep interest that allows to discover fundamental league-specific dynamics. In Table \ref{tab:theta}, we report for every league the estimates obtained at the end of the period considered and the related bootstrap $95\%$ confidence intervals.
The estimate for the Italian Serie A seems to indicate independent behaviours of home and away team performances. On the other hand, the Spanish La Liga is characterised by a positive value of $\hat{\theta}_3$. This evidence seems to confirm the hypothesis made in Section \ref{sec::exis_dep}, namely that the home (away) team tends to underperform when it plays against a team able to score less goals $y$ than a \textit{median} performance of the home (away) team (i.e., $F_{\lambda_k}(y)<0.5$).
Correspondingly, a home (away) team tends to overperform when it plays against a team able to score many goals (e.g., $F_{\lambda_k}(y)>0.5$), thus creating a more balanced match. Conversely, the opposite dynamics characterise the French Ligue 1, the German Bundesliga, and partially the English Premier League. 
Such dynamics seem to be mirrored in the common opinion according to which is easy to observe in the French Ligue 1 and the German Bundesliga more unbalanced matches than what expected by looking at the \textit{marginal} value of the teams.
In view of this, the ability to properly manage these different behaviours according to the specific league considered seems to represent the key feature of the Mar-Co model.

\begin{figure}
  \includegraphics[width=0.99\textwidth]{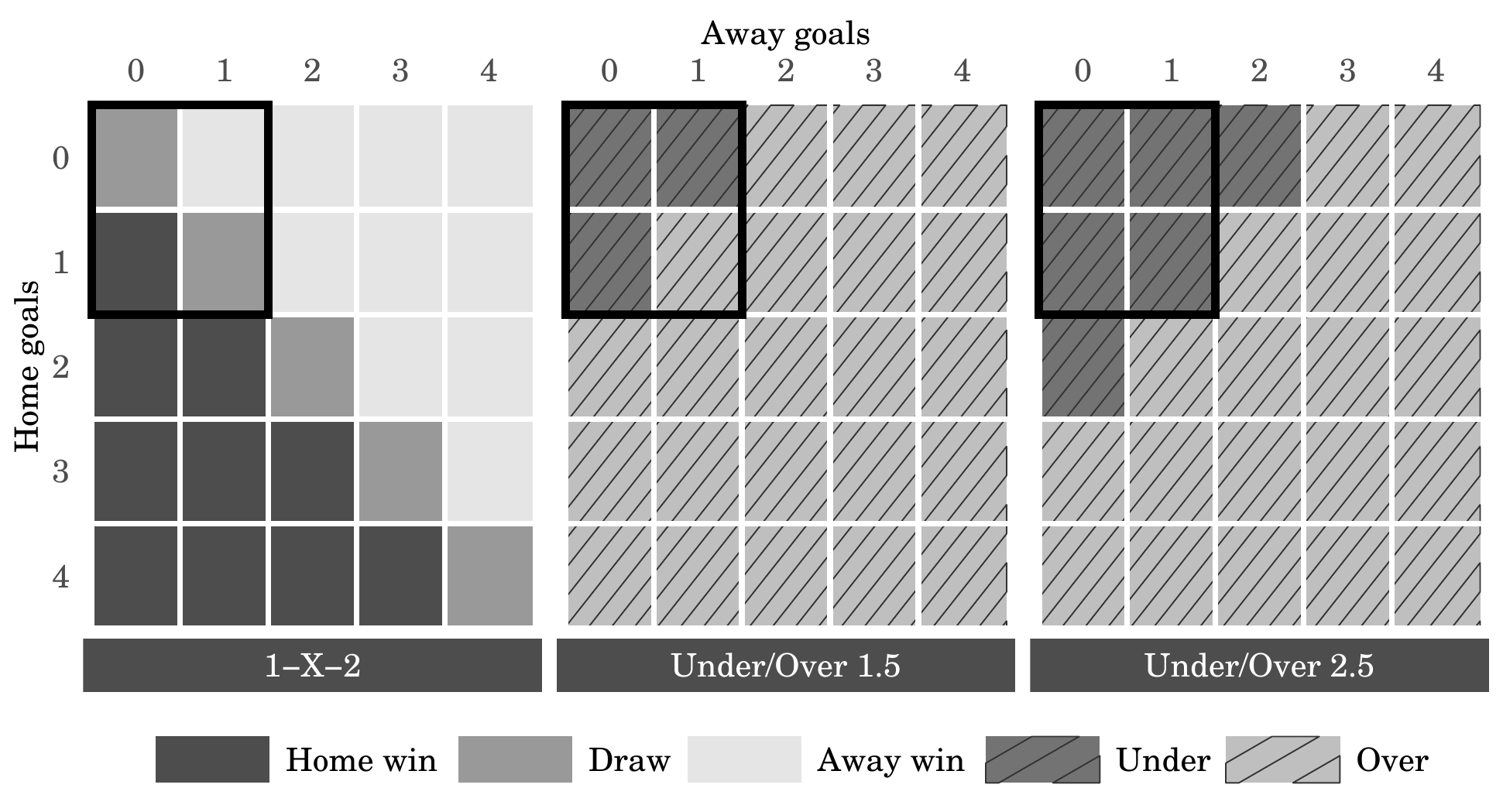}
\caption{Combinations of home and away goals characterising each betting outcome. The black boxes highlight the four match outcomes whose probabilities depend on $\rho$ in the Dixon and Coles model.}
\label{fig:bett}       
\end{figure}

Another betting type able to highlight the differences in terms of dependence modelling between the two approaches is the so-called Under/Over bet. In this case, the aim is to predict whether the overall number of goals scored during a game will be less (Under) or greater than (Over) a certain threshold. For the sake of clarity, Figure \ref{fig:bett}
shows the combinations of home and away goals characterising the two possible betting outcomes when the threshold is set equal to 1.5 (in the middle) and  2.5 (on the right). This type of bet is particularly interesting since, as noted in Section \ref{sec:dix_col_dep_str}, the Dixon and Coles parameter $\rho$ allows only the probabilities of the results (0,0), (1,0), (0,1), (1,1) to be reshuffled. Conversely, the newly introduced parameter $\theta_3$ allows also the probabilities of the other results to vary. 
As a result, assuming the other parameters as known, one is justified in expecting both dependence structures to have an impact on the prediction of 1-X-2 (see the left of Figure \ref{fig:bett}) and Under/Over 1.5, while only the dependence structure of the Mar-Co model is supposed to affect the prediction of Under/Over 2.5 since in that specific case the four cells contained in the black boxes in Figure \ref{fig:bett} belong to just one of the two possible events, i.e. Under 2.5.

\begin{figure}
  \includegraphics[width=0.99\textwidth]{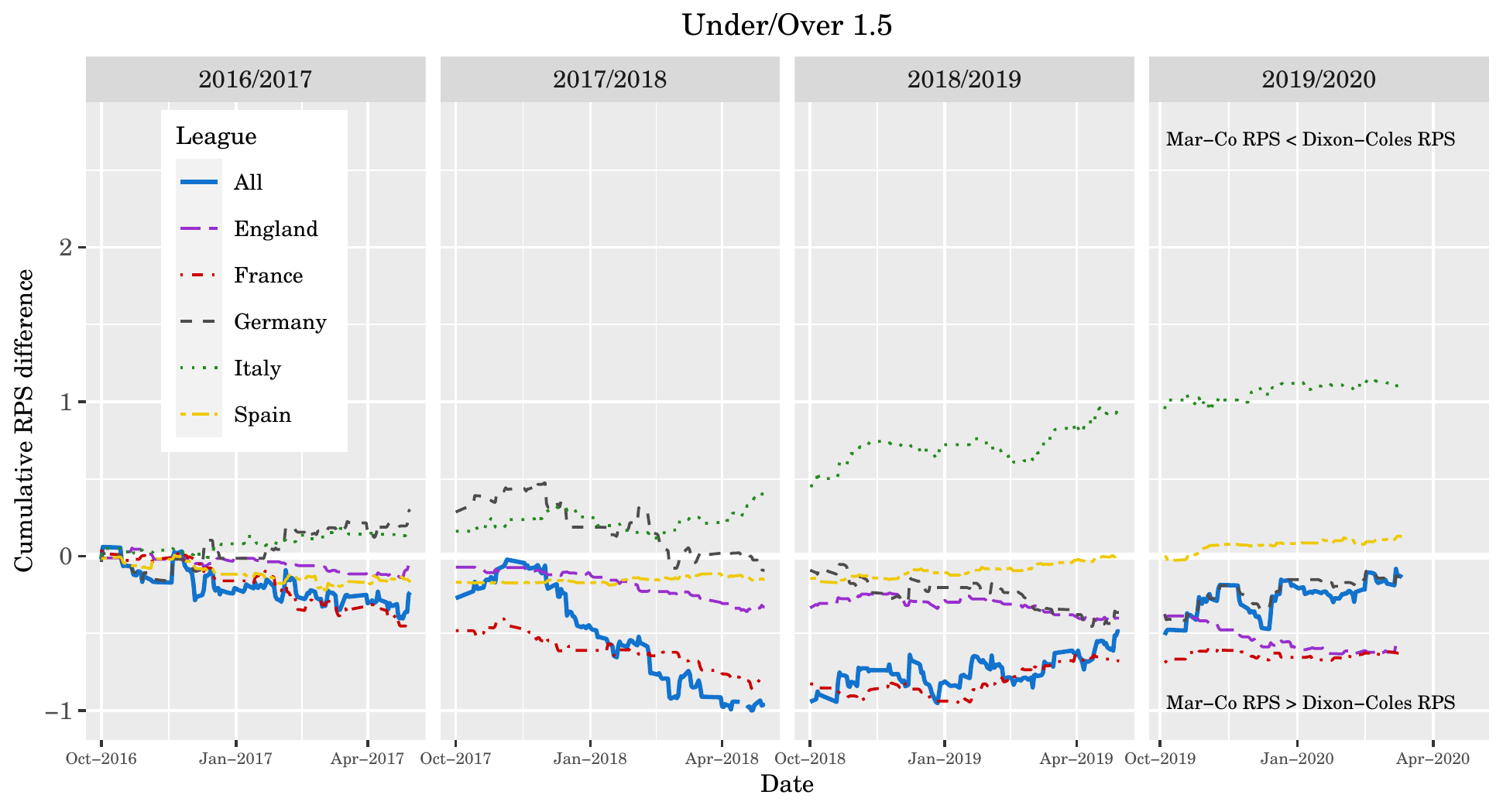}
\caption{Cumulative difference between the RPSs computed for the two competing models. Betting type: Under/Over 1.5.}
\label{fig:2prova}       
\end{figure}
\begin{figure}
  \includegraphics[width=0.99\textwidth]{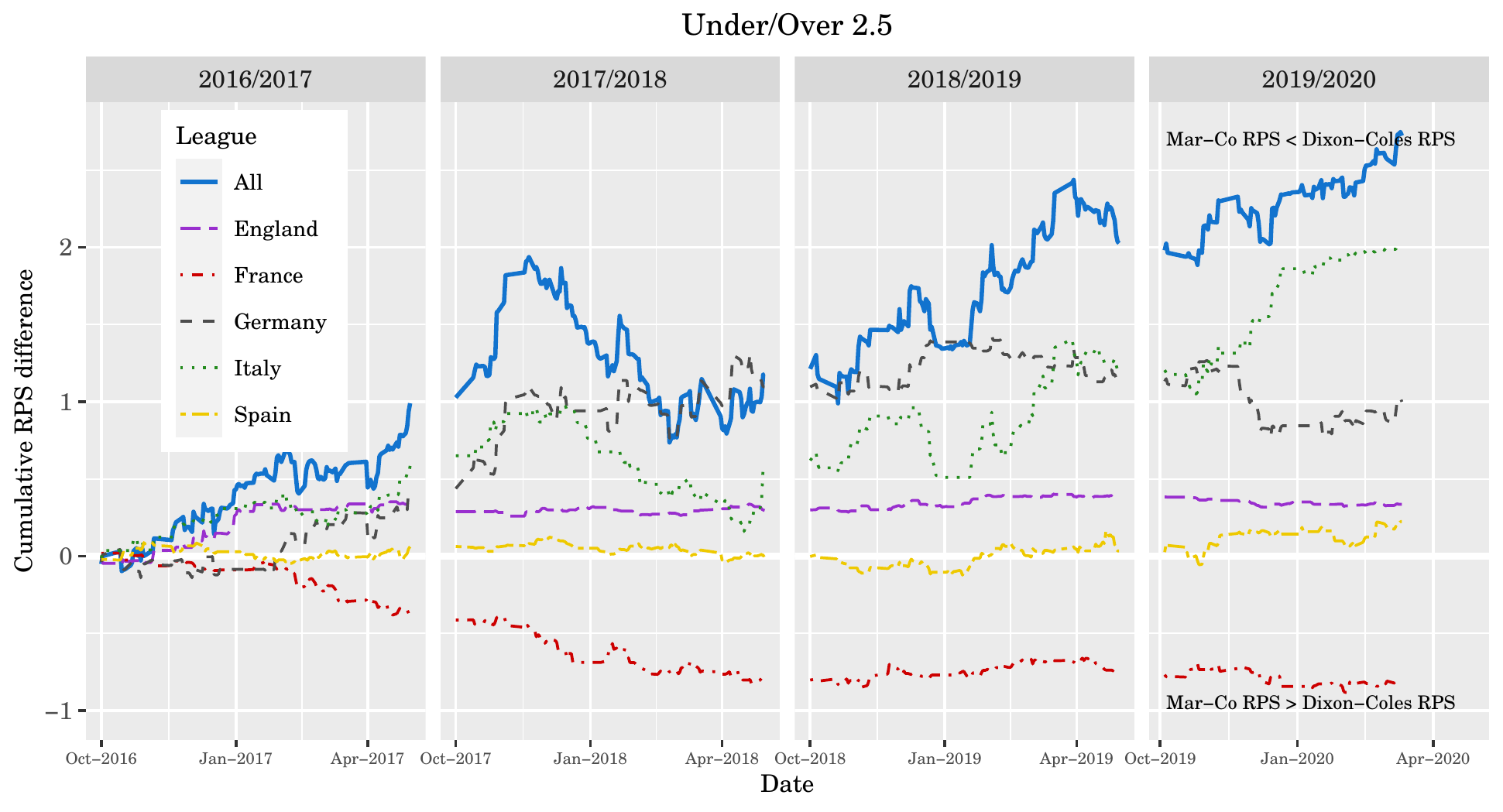}
\caption{Cumulative difference between the RPSs computed for the two competing models. Betting type: Under/Over 2.5.}
\label{fig:3}       
\end{figure}


Figure \ref{fig:2prova} and Figure \ref{fig:3} show the cumulative difference between the RPSs for Under/Over 1.5 and Under/Over 2.5, respectively. Focusing on Figure \ref{fig:2prova}, the Dixon and Coles model appears to outperform the Mar-Co one, with the sole exception of the Italian Serie A. However, the difference seems to be mainly due to chance: indeed, more than 40\% of the time the reshuffled samples obtained by replicating the aforementioned procedure provide a stronger evidence in favour of the Dixon and Coles model than the observed one. In view of this, the observed difference is not sufficient to draw conclusions about the Under/Over 1.5 framework.

As regards the Under/Over 2.5 betting type, Figure \ref{fig:3} highlights a totally different behaviour as the Mar-Co model largely outperforms the Dixon and Coles one, again with the sole exception of the French Ligue 1. 
The scale fixed scale of the y-axis in graphs \ref{fig:1}--\ref{fig:3} helps in showing the magnitude of the difference between the two models in this betting type, with respect to the others.
In this case, only $\sim$ 2\% of the time  the reshuffled samples  provide a stronger evidence in favour of the Mar-Co model than the observed one, and so the Mar-Co model seems to represent an interesting improvement of the Dixon and Coles model in the Under/Over 2.5 framework. This evidence is not entirely surprising given the specific betting type taken into account as any value of $\rho$ does not modify the probability of the event \emph{Under 2.5} (and consequently also of the event \emph{Over 2.5}) obtained with $\rho=0$. As a consequence, the fact that, given the estimates of the fixed effects and of $\gamma$, the estimates of the probabilities of Under 2.5 and Over 2.5 obtained by the Dixon and Coles model are equal to the estimates obtained assuming independence regardless the value of $\hat{\rho}$ represents ad undeniable limit in the Dixon and Coles procedure that may explain the improved performance obtained by the newly introduced Mar-Co model.

\section{Final remarks}
\label{sec::conclusion}
This work aimed to enrich the literature debate about goals-based models, by providing in-depth analyses on the distribution of the number of goals and by proposing an innovative adjustment of a classic and widely used model.

According to our contribution, nowadays the structure of the dependence between the number of goals scored by two competing teams in a match cannot be simplified to an adjustment of the probabilities of the outcomes 0-0, 1-0, 0-1, 1-1. This finding, jointly with the necessity of keeping a simple interpretation of the marginal distribution and their relation, motivates the alternative model proposed in this paper, which, moving from the Dixon and Coles approach, presents a more flexible and comprehensive representation of the dependence structure.
The particular specification of the term regulating the dependence, which is related to the conditional strength of the two competing teams, also plays a key role in providing interesting insights and interpretations of league dynamics.
Despite the lack of a full formalized protocol of comparison, the first encouraging results in terms of absolute predictive capability and further highlighted by the comparison between the Mar-Co model and the Dixon and Coles one confirm the validity of our proposal. This is particularly true in case of the Under/Over 2.5 bet, where the probability of the two betting outcomes, under the Dixon and Coles model, is not influenced by the dependence parameter.

However, some aspects must be pointed out. First of all, alternative types of bet (e.g. Asian Handicap bet) are worth exploring in order to obtain further evidences about the relation between dependence structure and prediction effectiveness.
Secondly, the inclusion of external covariates, such as information on the players conditions or on the teams motivations, should be considered to make the model even more attractive for an effective use in the world of betting.
Finally, the lack of a recursive estimation process represents an undeniable limit in terms of computational effort required, since new crucial information is provided after each match.

However, although an extended analysis on a longer period and on more leagues would be desirable, the promising results achieved shed new light on the importance of properly choosing a suitable dependence structure to model the number of goals in a football match.

\section*{Declarations}

\subsection*{Funding} Not applicable.

\subsection*{Conflicts of interest} The authors declare that they have no conflict of interest.

\subsection*{Availability of data and material} Data are publicly available at \url{http://www.football-data.co.uk/}.

\subsection*{Code availability} The {\sf R} code used in this study is available from the corresponding author upon reasonable request.

%
%

\bibliographystyle{apa}
\bibliography{bib_file}   

%
%

\end{document}